\documentclass[11pt]{article}
\usepackage{amsmath,amssymb}
\usepackage{epsfig}
\begin{document}
%\begin{titlepage}
\begin{flushright} SU-4252-783\\
SINP/TNP/03-19
             \\
\end{flushright}         
\begin{flushright}   July 2003
             \\
\end{flushright}   
\begin{center}
   \vskip 3em
{\LARGE EDGE CURRENTS IN NON-COMMUTATIVE \vskip 1em CHERN-SIMONS THEORY
FROM A NEW MATRIX MODEL}
  \vskip 2em
{\large A. P. Balachandran$^\dagger$, Kumar. S. Gupta$^{*}$
\footnote[1]{Regular Associate, The Abdus Salam ICTP, Trieste, Italy}
and
S. K\"{u}rk\c{c}\"{u}o\v{g}lu$^\dagger$
\footnote[2]{E-mails: bal@phy.syr.edu, gupta@theory.saha.ernet.in, 
skurkcuo@phy.syr.edu,} 
\\[2em]}
\em{\oddsidemargin 0 mm
$^\dagger$Department of Physics, Syracuse University,
Syracuse, NY 13244-1130, USA\\
\vskip 1em
$^*$ Theory Division, Saha Institute of Nuclear Physics,\\
1/AF Bidhannagar, Calcutta - 700064, India}
\end{center}
\vskip 1em
\begin{abstract}

This paper discusses the formulation of the non-commutative Chern-Simons
(CS) theory where the spatial slice, an infinite strip, is a manifold with
boundaries. As standard $*$-products are not correct for such manifolds, the
standard non-commutative CS theory is not also appropriate here. Instead we
formulate a new finite-dimensional matrix CS model as an approximation to
the CS theory on the strip. A work which has points of contact with ours is
due to Lizzi, Vitale and Zampini where the authors obtain a description for
the fuzzy disc.  The gauge fields in our approach are operators supported on
a subspace of finite dimension $N + \eta$ of the Hilbert space of
eigenstates of a simple harmonic oscillator with $N\,, \eta \in {\mathbb
Z}^+$ and $N \neq 0$.  This oscillator is associated with the underlying
Moyal plane. The resultant matrix CS model has a fuzzy edge. It becomes the
required sharp edge when
$N$ and $\eta \rightarrow \infty$ in a suitable sense. 
The non-commutative CS theory on the strip is defined by this limiting
procedure. After performing the canonical constraint analysis of the matrix
theory, we find that there are edge observables in the theory generating a
Lie algebra with properties similar to that of a non-abelian Kac-Moody
algebra. Our study shows that there are $(\eta +1)^2$ abelian
charges(observables) given by the matrix elements $({\hat{\cal
A}}_i)_{N-1\,N-1}$ and
$({\hat{\cal A}}_i)_{nm}$ (where $n\,or\, m \geq N$) of the gauge fields,
that obey certain standard canonical commutation relations. In addition, the
theory contains three unique non-abelian charges, localized near the
$N^{th}$ level. We observe that all non-abelian edge observables except
these three can be constructed from the $ (\eta+1)^2$ abelian charges above. 
Using some of the results of this analysis we discuss in detail the limit
where this matrix model approximates the CS theory on the infinite strip. 
Finally, we include a short section containing our comments on the
commutative limit of our model, where we also give a closed formula for the
central charge of the Kac-Moody-like algebra of the non-commutative CS
theory in terms of the diagonal coherent state matrix elements of operators
and star products.
\end{abstract}

\newpage

\setcounter{footnote}{0}

\section{Introduction}

Recently, there has been much interest in formulating Chern-Simons theories
on non-commutative manifolds \cite{Kluson:2000mv}.  In \cite{susskind} the
non-commutative CS theory has been used for the purpose of obtaining a new
description of the Quantum Hall Effect (QHE). Polychronakos \cite{poly} and
Morariu and Polychronakos \cite{mor} have proposed finite-dimensional matrix
models of CS theory and used them to explain certain properties of the
Fractional Quantum Hall Effect (FQHE).

In this paper we report on our work on the formulation of non-commutative CS
theory where the spatial slice, an infinite strip, is a manifold with
boundaries. It can serve to describe non-commutative QHE in such a strip.

There have been previous attempts to carry out such a formulation on a disc
and on a half plane in the presence of spatial non-commutativity
\cite{pinzul1, lugo}. The main obstacle faced in these attempts can be
traced to the absence of a well-defined $\star$-product for these types of
manifolds. We discuss this problem in the beginning of section 3 in some
detail to better explain our motivation and the need for the matrix CS model
we introduce afterwards. Recently, Pinzul and Stern \cite{pinzul2} have
studied the CS theory written on a non-commutative plane with a `hole'. 
They have shown that in this case the algebra of observables is a nonlinear
deformation of the $w_\infty$ algebra. In another work \cite{lizzi2}, Lizzi,
Vitale and Zampini have formulated a fuzzy disc, on which the
non-commutative CS theory can be formulated.  It has overlaps with our work
here and has also points of contact with \cite{pinzul1}.

Our model stems from taking gauge fields as operators supported on an $N+
\eta$-dimensional subspace of the Hilbert space,
where $N \,, \eta \in {\mathbb Z}^+$ and $N \neq 0$.  It is spanned by the
eigenstates of a simple harmonic oscillator associated to the underlying
Moyal plane. The resultant matrix CS model has a fuzzy edge and it becomes
the required sharp edge when $N$ and $\eta \rightarrow \infty$ in a suitable
sense. The non-commutative CS theory on the strip is defined by this
limiting procedure. In this framework we perform the canonical constraint
analysis for the finite-dimensional matrix CS model and find out that there
are edge observables in the theory generating a Lie algebra with properties
similar to that of a non-abelian Kac-Moody algebra. Our study shows that
there are $(\eta +1)^2$ ``abelian'' charges (observables) given by the
matrix elements
$({\hat{\cal A}}_i)_{N-1 \,N-1}$ and
$({\hat{\cal A}}_i)_{nm}$ (where $n\,or\, m \geq N$) 
of the gauge fields, that obey certain standard canonical commutation
relations. In addition, the theory contains three unique ``non-abelian''
charges, localized near the $N^{th}$ level. We show that all ``non-abelian''
edge observables, except these three can be constructed from the $
(\eta+1)^2$ ``abelian'' charges above.  Using some of the results of this
analysis we discuss in detail the limit where this matrix model approximates
the CS theory on the infinite strip. Finally, we discuss the commutative
limit of our model, where we also give a closed formula for the central
charge of the Kac-Moody-like algebra of the non-commutative CS theory in
terms of the diagonal coherent state matrix elements of operators and star
products.

The organization of our paper is as follows: In section 2 we start with a
brief review of the formulation of edge currents in a commutative CS theory
on an infinite strip. Here we also demonstrate how one can formulate these
edge effects while working on the full plane and encoding the boundary
structure in the test functions. The next three sections systematically
develop the ideas outlined in the last paragraph.  We conclude with
highlighting the main results of our work and their possible physical
significance.

\section{Edge Currents Revisited}

Here we briefly review the derivation of Kac-Moody algebra obeyed by edge
observables in an abelian Chern-Simons theory. Results summarized in this
section are well-established and extensively discussed in
\cite{Witten:1988hf,bal1}. 
We consider the simplest case of an infinite strip (say in $x_2$ direction)
in ${\mathbb R}^2 \otimes {\mathbb R}^1$, where ${\mathbb R}^1$ stands for
time.  This formulation closely follows the one given in \cite{bal1} for a
disc. Here we also describe another way in which we can work on the full
plane and encode the boundary structure in the domains of the test
functions. This will serve as a prototype approach we shall use for the
finite-dimensional matrix CS model.

Let $x \equiv (x_1, x_2)$ denote the points of ${\mathbb R}^2$.  The strip
${\cal T}$ is defined by
\begin{equation}
{\cal T}= \{ (x_1,x_2) \in {\mathbb R}^2 \big| -L \leq x_1 \leq L \} \,.
\end{equation}

The usual action for the abelian Chern-Simons theory on ${\cal T}$ is
\begin{equation}
S= \frac{\kappa}{4 \pi} \int_{{\cal T} \otimes {\mathbb R}^1} A 
\wedge dA \quad \quad \quad A = A_\mu dx^\mu \,.
\end{equation}
The gauge fields satisfy the equal time Poisson Brackets (P.B.'s)
\begin{equation}
\{ A_i(x_0, x_1, x_2) \,, A_j(x_0, x_1^\prime \,, x_2^\prime) \}= 
\varepsilon_{ij} \frac{2\pi}{\kappa} 
\delta^2(\overrightarrow{x}- \overrightarrow{x}^\prime) \,,
\label{eq:pb1} 
\end{equation}
where $i,j = 1,2$ and $\varepsilon_{12} = -\varepsilon_{21} =1$.

From the variation of $S$ under $A_0$ we have the 
Gauss law constraint, which can be written by
introducing test functions $\Lambda^0$ as
\begin{equation}
g(\Lambda^0)= \frac{\kappa}{2 \pi} \int_{\cal T} 
\Lambda^0 (x)dA(x) \approx 0 \,.
\label{eq:gauss1}
\end{equation}
Differentiability of $g(\Lambda^0)$ imposes the condition $\Lambda^0
{\big|}_{\partial {\cal T}}= 0$.  Using (\ref{eq:pb1}) and (\ref{eq:gauss1})
it can be shown that $g(\Lambda^0)$ are first class constraints:
\begin{equation}
\{ g(\Lambda_1^0) \,, g(\Lambda_2^0) \} \approx 0 \,,
\label{eq:can1}
\end{equation}
and generate the gauge transformations $A \rightarrow A - d \Lambda^0$.

The charges of this theory are 
\begin{equation}
q(\Lambda) =\frac{\kappa}{2\pi} \int_{\cal T} d\Lambda \wedge A \,.
\label{eq:can2}
\end{equation}
Since they are first class:
\begin{equation}
\{ q(\Lambda) \,, g(\Lambda^0) \} \approx 0 \,,
\label{eq:can3}
\end{equation} 
they constitute the observables of this theory. 
We note that $\Lambda$ are test functions associated with
$q(\Lambda)$'s and unlike $\Lambda^0$ they in general do 
not vanish on $\partial{\cal T}$.

Now note that for $\Lambda_1- \Lambda_2 =\Lambda^0$, 
\begin{equation}
q(\Lambda_1)- q(\Lambda_2) = - g(\Lambda_1 - \Lambda_2) \approx 0 \,. 
\label{eq:edgeo1}
\end{equation}
This means that test functions  which are equal on $\partial{\cal T}$
generate charges which are equal modulo constraints.  Thus, $q(\Lambda)$ are
indeed edge observables. They generate the transformations
$A \rightarrow A - d \Lambda$ which do not necessarily vanish on 
$\partial {\cal T}$.     

The P.B.'s of charges can be identified to be the 
$U(1)$ Kac-Moody algebra on $\partial {\cal T}$, since
\begin{equation}
\{ q(\Lambda_1), q(\Lambda_2) \} = 
\frac{\kappa}{2 \pi} \int_{\cal T} d \Lambda_1 \, \wedge \,d \Lambda_2
=\frac{\kappa}{2 \pi} \int_{\partial {\cal T}} \Lambda_1 \,d \Lambda_2 \,.
\label{eq:kac1}
\end{equation}
Choosing the basis for the test functions on $\partial {\cal T}$ as
\begin{equation}
\Lambda_1 \big|_{x_1 = L} = e^{i \,k_1 \, x_2} \,, \quad \quad 
\Lambda_2 \big|_{x_1 = L} = e^{i \,k_2 \, x_2} \,, \quad \quad
\Lambda_1 \big|_{x_1 = -L} = 0 \,, \quad \quad 
\Lambda_2 \big|_{x_1 = -L} = 0 \,,
\label{eq:boundary1}
\end{equation}
(\ref{eq:kac1}) can be written as
\begin{equation}
\{ q(\Lambda_1), q(\Lambda_2) \} = i\kappa \,k_2 \delta(k_1+k_2) \,,
\label{eq:kac2}
\end{equation}
which is the usual form of the $U(1)$ Kac-Moody algebra.

An equivalent formulation of the above can be given as follows. 
Consider the CS theory on ${\mathbb R}^2 \otimes {\mathbb R}^1$. 
However, now the spatial components of the gauge fields are 
supported in the region $|x_1| \leq L$ (i.e. in ${\cal T}$), whereas
$A_0$ has support in $|x_1| < L$.  
Obviously, (\ref{eq:pb1}) holds now only if $x_1 \,,x_1^\prime 
\in {\cal T}$, otherwise its R.H.S. vanishes. 

From the variation of $S$ under $A_0$ we still have the Gauss law constraint 
\begin{equation}
G(\Lambda^0) = \frac{\kappa}{2 \pi} 
\int_{{\mathbb R}^2} \Lambda^0 dA \approx 0,
\end{equation}
where now the integration is over ${\mathbb R}^2$. 
The condition that $\Lambda^0$'s are supported in $|x_1| < L$
follows from that of $ \delta A_0$ which is of course the same 
as that of $A_0$. Thus we have  
\begin{equation}
\Lambda^0 = 0 \quad \quad \quad for \quad \quad |x_1| \geq L.
\label{eq:can4}
\end{equation}

The results of the canonical analysis 
(given in (\ref {eq:can1})-(\ref {eq:can3})) go through where now 
integrals are over all ${\mathbb R}^2$ and $\Lambda^0$ is as given 
in (\ref{eq:can4}). This establishes
$G(\Lambda^0)$'s as first class constraints and 
\begin{equation}
Q(\Lambda) = \frac{\kappa}{2 \pi} \int_{{\mathbb R}^2}  
d \Lambda \wedge A  
\label{eq:cano1}
\end{equation}
as the observables. Clearly, $\Lambda$ can be supported 
on all of ${\mathbb R}^2$, but if it is
supported only on ${\mathbb R}^2 \setminus {\cal T}$ (i.e. $|x_1| > L$) 
then we immediately see from (\ref{eq:cano1})
that $Q(\Lambda) \equiv 0$.  

For $\Lambda_1 - \Lambda_2 = \Lambda^0$ we have
\begin{equation}  
Q(\Lambda_1)- Q(\Lambda_2) = - G(\Lambda_1 - \Lambda_2) \approx 0 \,. 
\end{equation}
It follows from this and the remark after 
(\ref {eq:cano1}) that $Q(\Lambda)$ are observables
localized at $x_1 = \pm L$.
The P.B.'s of charges gives the Kac-Moody algebra:
\begin{equation}
\{ Q(\Lambda_1), Q(\Lambda_2) \} = 
\frac{\kappa}{2 \pi} \int_{{\mathbb R}^2} d \Lambda_1
\, \wedge \,d \Lambda_2 \,.
\label{eq:kak}
\end{equation}

A suitable choice for $\Lambda$'s to compute this algebra at $x_1 = L$ is 
\begin{equation}
\Lambda_i = \theta(x_1 - L)\, e^{i\, k_i \,x_2} \,, \quad (i = 1,2) \,,
\label{eq:kak1}
\end{equation}
where $\theta(x_1 - L)$ is the step function centered 
at $x_1=L$ with $\theta(0) = 1$. 
Then we have from (\ref{eq:kak}) that
\begin{equation}
\{ Q(\Lambda_1), Q(\Lambda_2) \} = i\kappa\,k_2 \delta(k_1+k_2) \,,
\label{eq:kac3}
\end{equation}
which is the same as (\ref{eq:kac2}). In order to get the 
algebra of observables 
at $x_1 = - L$ one has to replace (\ref{eq:kak1}) by
\begin{equation}
\Lambda_i = \big(1- \theta(x_1 +L) \big) \, e^{i\, k_i \,x_2} \,. 
\end{equation}
This way of treating edge properties is not 
completely new and has examples in planar systems of
condensed matter physics \cite{ezawa}. We now turn our 
attention to the treatment
of the non-commutative case.

\section{Non-commutative Chern-Simons Theory on the Infinite Strip}

\subsection{Remarks on the Non-commutative 
CS theory on a Manifold with Boundaries } 

As pointed out by \cite{pinzul1, lugo} the canonical formulation of 
Chern-Simons theory on a non-commutative
manifold with boundaries (such as a disc, 
an infinite strip etc.) presents serious difficulties. In fact the
problem underlying such a formulation is that the standard $\star$-product 
on the Moyal plane is no longer
valid on manifolds with boundaries. To make this point 
clear and supply enough mathematical
basis for this assertion let us consider first the Chern-Simons 
action on the Moyal plane with no boundaries:
\begin{equation}
S_{NCCS}= - \frac{\kappa}{4\pi} \int dx_0 \,d^2 x \,
\varepsilon_{\mu \nu \lambda}\,
\big( A_\mu \star_M \partial_\nu A_\lambda + 
\frac{2}{3} \,A_\mu \star_M A_\nu \star_M
A_\lambda \big) \,.
\label{eq:nccs}
\end{equation}  
Here $(\mu, \nu, \lambda = 0,1,2)$, $x_0$ is time
$x_1$ and $x_2$ are coordinates on
the Moyal plane and $\varepsilon_{123}=1$.

The Moyal algebra is characterized by the $\star$-product
\begin{equation}
f \star_M g (x_1, x_2) = f(x_1, x_2) e^{\frac{i \theta}{2} 
( {\overleftarrow \partial}_{x_1} \,
{\overrightarrow \partial}_{x_2} - {\overleftarrow \partial}_{x_2} \,
{\overrightarrow \partial}_{x_1})} g(x_1, x_2) \,.
\label{eq:moyal1}
\end{equation}
Defining the $\star$-commutator of $f$ and $g$ by
\begin{equation}
\lbrack f, g\rbrack_{\star_M} = f \star_M g - g \star_M f \,,
\end{equation}
the spatial non-commutativity can be expressed as 
\begin{equation}
\lbrack x_1 \,, x_2 \rbrack_{\star_M} = i \theta \,,
\end{equation}
$\theta$ being the non-commutativity parameter.

Naively, one may consider $S_{NCCS}$ on a manifold 
with boundaries, say the infinite strip ${\cal T}$, and write the
Gauss law constraint as 
\begin{equation}
g(\Lambda^0) = - 
\frac{\kappa}{2 \pi} \int_{{\cal T}} d^2 \,x \,
\varepsilon_{ij}\, \Lambda^0 \, 
(\partial_i \,A_j + A_i \star_M A_j) \approx 0 \,,
\end{equation} 
where $(i,j = 1,2)$ and $\Lambda^0 \big|_{\partial {\cal T}} = 0$.
However, this expression as well as the $S_{NCCS}$ when written on 
the strip  ${\cal T}$
is not well defined. 
%However, now the algebra of constraints do not close, that is 
%\begin{equation}
%\{ g(\Lambda_1^0) \,, g(\Lambda_2^0) \}
%\label{eq:gn1} 
%\end{equation}
%is not weakly zero. 
This is because $\star_M$ product does not exist on ${\cal T}$.
To prove this fact we note that the formula for the $\star_M$ 
product in (\ref{eq:moyal1}) contains the 
exponential of the differential operator $-i \partial_{x_1}$. 
With the usual definition of its domain $-i \partial_{x_1}$  
generates translations
so that $e^{i(-i c \partial_{x_1})}$ translates functions of $x_1$ by $c$ :
$\big(e^{i(-i c \partial_{x_1})} \big) \psi (x_1) = \psi (x_1 + c)$. 
Consequently, if $\psi$ has support
$[-L, L]$ , $e^{i(-i c \partial_{x_1})} \psi$ does not,  
and $\star_M$ is not defined on
functions supported in $[-L, L]$ \cite{simon}.
%This explains why
%a P.B. like the one given in (\ref{eq:gn1}) is not weakly zero \cite{simon}.

Alternatively, to circumvent the 
impossible task of obtaining a well-defined and useful
$\star$-product on these types of manifolds we propose a 
finite-dimensional matrix model where
the edges are fuzzy. It becomes the 
CS theory on a non-commutative infinite strip in the limit
where the size of the matrices approaches infinity. In this 
respect, we emphasize that our approach is
completely different from the previous attempts 
in the literature as the following section illustrates. 

\subsection{The Matrix Model}
We now describe our matrix model. 
Since we will be working in the operator formalism in this subsection,
we discriminate the operators from the elements of the corresponding 
Moyal $\star$ algebra of functions
by putting a hat symbol on the elements of the former. 
Thus the Moyal plane is described by operators
${\hat x_i} \, (i =1,2)$ with the relation  
\begin{equation}
\lbrack {\hat x}_1, {\hat x}_2 \rbrack = i\theta \,.
\end{equation}

To supply the mathematical basis for our arguments 
we think of a simple harmonic oscillator in
the $x_1$-direction, described by the Hamiltonian
\begin{equation}
{\hat H}= \frac{{\hat x}_2^2}{2m}+ \frac{1}{2}k {\hat x}_1^2 \,,
\end{equation}
and the oscillation frequency $\omega=\sqrt{\frac{k}{m}}$.
Now consider the Hilbert space ${\cal H}$ spanned by the 
eigenstates of this Hamiltonian.
In the matrix CS model we will 
construct ${\cal H}$ will be associated to the underlying Moyal plane,
and its finite-dimensional subspaces will 
serve as the carrier spaces
of the operators that we are 
going to introduce in our matrix model. More precisely, we will define
these operators in terms of their action on the elements of a 
finite-dimensional subspace of 
${\cal H}$ and its orthogonal complement.

The number of energy eigenstates of this Hamiltonian below the energy
$E = \frac{1}{2} k L^2$ with $L$ being the maximum classical amplitude 
(given by the location of the edges)
is finite and given by   
\begin{equation}
M= \big\lbrack \frac{k L^2+ \theta \omega }{2 \theta \omega} \big\rbrack ,
\label{eq:number1}
\end{equation}
where $\lbrack \frac{k L^2+ \theta \omega}{2 \theta \omega} \rbrack$ 
is the largest integer smaller
than $\frac{k L^2+ \theta \omega}{2 \theta \omega}$.
These $M$ states span a subspace ${\cal H}_M$ of the harmonic oscillator 
Hilbert space
${\cal H}$ and they can be taken as an orthonormal 
basis in ${\cal H}_M$.\footnote{To avoid any
confusion that our notation might create later on we note that 
the Fock basis we are
using for ${\cal H}$ is the usual one 
where states are labeled starting with quantum 
number $0$, therefore the top state in ${\cal H}_M$ has 
the quantum number $M-1$.}  
From (\ref{eq:number1}) it is easy to see that
keeping both oscillation frequency $\omega$ and
the maximum oscillation amplitude $L$ fixed while 
increasing $k$ (i.e. steepening the potential)
results in larger number of states with energy $E \leq \frac{1}{2} k L^2$. 
Hereafter we keep $\omega$
fixed unless otherwise stated. Let us consider a level
$K$ with $K \leq M$. Then the large $K \approx M$ levels 
with energy $E \approx \frac{1}{2} k L^2$
get localized near the edges $x_1=\pm L$, 
since their probability amplitudes are maximum there.
On the contrary, those levels with $E \ll \frac{1}{2} k L^2$ are 
localized well inside
$-L \leq x_1 \leq L$ for large $M$. We also note that all levels 
decay exponentially outside
$|x_1| \leq L$. Linear operators on ${\cal H}_M$ which are 
zero on the orthogonal complement ${\cal H}_M^\perp$ 
behave also in a similar fashion. 
Later we will show that diagonal coherent state matrix elements of 
operators, say like $|K><K|$ with $K \approx M$
will peak near $\pm L$ for all $|x_2|<M$ for large $M$, while those with $K$ 
much less than $M$ acquire maxima within
$|x_1|<L$. The characteristic width of the peaks at $\pm L$, which 
gets narrower as $M$ gets larger,
gives us a natural scale to which we will relate the 
thickness $\Delta \ell$ of the
``fuzzy edge'' of our matrix model.

For the finite-dimensional matrix CS theory we have in mind, we treat spatial
components of the gauge fields as operators supported on this $M$-dimensional 
subspace ${\cal H}_M$ of
${\cal H}$. This and the 
physical picture described above for the 
underlying Moyal plane imply two immediate
consequences:
\begin{itemize}
\item It is those states contained within 
the thickness $\Delta \ell$ of the edges that will be responsible
for the edge observables in our finite-dimensional matrix CS model;
\item The $M \rightarrow \infty$ limit of this matrix CS theory can 
be taken to
define the non-commutative CS theory on the infinite 
strip with boundaries at $x_1 = \pm L$. 
\end{itemize}
After writing the matrix CS model and performing the constraint analysis for
it, we will find these observables, and we will have the necessary
information to explain and elaborate on the properties of the large $M$
limit and how it is realized.

We now split $M$ as $M= N+\eta$ where $N\,, \eta \in {\mathbb Z}^+$ 
and $N \neq 0$, 
and use this separation of the total dimension to 
define the domains and the ranges of
our operators. In our formalism the gauge fields ${\hat A}_\mu\,, 
(\mu= 0,i~ {\rm where}~ i= 1,2)$ are
anti-Hermitian operators that we choose to take in the way given below:
\begin{eqnarray}
{\hat A}_i {\cal H}_{N + \eta} &\subseteq& {\cal H}_{N + \eta}\,, \quad \quad 
{\hat A}_i {\cal H}_{N + \eta}^\perp = \{0\} \,, \nonumber \\
{\hat A}_0 {\cal H}_{N-1} &\subseteq& {\cal H}_{N-1}\,, \quad \quad
{\hat A}_0 {\cal H}_{N-1}^\perp = \{0\} \,.  
\label{eq:gaugef1}
\end{eqnarray}
Here for any $K \in {\mathbb Z}^+$,${\cal H}_{K}^\perp$ denotes the 
orthogonal complement of ${\cal H}_{K}$.
The fact that ${\hat A}_0$ is nonzero only on ${\cal H}_{N-1}$ and not on
${\cal H}_{N+\eta}$ should be noted. The reason behind this condition will 
be explained later
in connection with the canonical analysis of the model.

If $|n>$ denotes the $n^{th}$ normalized energy level 
and ${\hat P}_{nm}$ is the operator given by
${\hat P}_{nm}=|n><m|$, then we can write,
\begin{equation}
{\hat A}_i = \sum_{n,m=0}^{(N-1 + \eta)} i 
({\hat {\cal A}}_i)_{nm} \,{\hat P}_{nm} \,, \quad 
({\hat {\cal A}}_i)_{nm} = 0  \quad for \quad n \quad 
or \quad m > N-1 + \eta \,.
\label{eq:gaugea}
\end{equation}
For ${\hat A}_0$, the same equation is 
valid if we replace $N-1+\eta $ with $N-2$.
It also follows from ${\hat A}_\mu^\dagger = - {\hat A}_\mu$ that
\begin{equation}
({\hat {\cal A}}_\mu)_{nm}^* = - ({\hat {\cal A}}_\mu)_{mn} \,.
\end{equation}

There is another way to express (\ref{eq:gaugef1}). For this 
let us introduce the orthogonal projector  ${\bf \hat{1}}_{K}$ by
\begin{eqnarray}
&&{\bf \hat{1}}_{K} = 
\sum_{n\,, m=0}^{K-1} ({\bf \hat{1}}_K)_{nm} {\hat P}_{nm}
:= \sum_{n\,, m=0}^{K-1} (\delta)_{nm} {\hat P}_{nm} = \sum_{n=0}^{K-1} 
{\hat P}_{nn} \,, \nonumber \\
&&({\bf \hat{1}}_K)_{nm} = (\delta)_{nm} = 
0 \quad for \quad n \quad  or \quad m > K-1 \,.
\label{eq:orthgp1}
\end{eqnarray} 

Then (\ref{eq:gaugef1}) is equivalent to 
\begin{equation}
{\hat A}_i = 
{\bf \hat{1}}_{N+\eta} \,{\hat A}_i \,{\bf \hat{1}}_{N+\eta} \,,\quad \quad    
{\hat A}_0 = {\bf \hat{1}}_{N-1} \,{\hat A}_0 \,{\bf \hat{1}}_{N-1} \,.
\label{eq:orthgp2} 
\end{equation}

The Chern-Simons Lagrangian for the model 
reads (up to a total space derivative)
\begin{equation}
L_{NCCS} = -\frac{\kappa \theta}{2} 
\varepsilon_{ij} \,Tr \,\big( -{\hat A}_i {\hat {\dot A}}_j
+ 2 {\hat A}_0 (\partial_i {\hat A}_j + {\hat A}_i {\hat A}_j) \big) \,,
\end{equation} 
where ${\hat {\dot A}}_j = \partial_0 {\hat A}_j$, 
trace is over the Hilbert space ${\cal H}$, and derivations are given by
\begin{equation}
\partial_i (.)  
= \frac{i}{\theta} \,\varepsilon_{ij} \lbrack{\hat x}_j \,, (.) \rbrack \,.
\label{eq:derv}
\end{equation}

Several remarks about properties of $L_{NCCS}$ are in order. 
First, $L_{NCCS}$ changes by total derivatives
under infinitesimal gauge 
transformations of the form  
\begin{equation}
{\hat A}_\mu \rightarrow {\hat A}_\mu + 
\big(\partial_\mu {\hat \lambda} + i\lbrack {\hat A}_\mu \,,
{\hat \lambda} \rbrack \big) \,,
\label{eq:gt}
\end{equation}
where ${\hat \lambda}$ is a matrix with infinitesimal elements.
Next, as in Chern-Simons theory on a commutative manifold, 
the conjugate momenta ${\hat \Pi}_0$ to
${\hat A}_0$ are weakly equal to zero and first class, 
thus ${\hat A}_0$ is not an observable and
can be eliminated from the rest of our discussion.

The equal time P.B.s of $({\hat {\cal A}}_i)_{nm}$ can be written as
\begin{equation}
\big\{({\hat {\cal A}}_i)_{nm} \,, ({\hat {\cal A}}_j)_{rs} \big\} 
= \frac{1}{\kappa \theta} \varepsilon_{ij} ({\bf \hat{1}}_{N+\eta})_{ns} 
({\bf \hat{1}}_{N+\eta})_{mr} 
= \frac{1}{\kappa \theta} \varepsilon_{ij} \delta_{ns} \delta_{mr}
\,, \quad \quad \quad  n,m,r,s \in \lbrack 0, N-1+\eta \rbrack \,, 
\label{eq:canonical2}
\end{equation}
which in the operator formalism is the statement 
that $({\hat {\cal A}}_1)_{nm}$ and
$({\hat {\cal A}}_2)_{rs}$ are canonically conjugate. In terms of 
the operators
${\hat A}_i$ in (\ref{eq:gaugea}), this is
\begin{equation}
\big\{{\hat A}_i\,, {\hat A}_j \big\} = - \frac{1}{\kappa 
\theta} (N+\eta)\, \varepsilon_{ij} \,
\sum_{n=0}^{N-1+\eta} {\hat P}_{nn} = - \frac{1}{\kappa 
\theta} (N+\eta) \varepsilon_{ij}
\,{\bf \hat{1}}_{N+\eta} \,.
\label{eq:pro1}
\end{equation}
We now turn our attention to the canonical constraint analysis of this model.

\subsection{Canonical Analysis of the Matrix Model}

From the variation of $L_{NCCS}$ with 
respect to ${\hat A_0}$, we have the Gauss law constraint
\begin{equation}
- \kappa \,\theta \,\varepsilon_{ij} \,Tr\,\big(\delta {\hat A}_0
(\partial_i {\hat A}_j + {\hat A}_i {\hat A}_j) \big) \approx 0 \,.
\label{eq:A0}
\end{equation} 
As $\delta {\hat A}_0$ is not zero only in ${\cal H}_{N-1}$, we find
\begin{equation}
g({\hat \Lambda}^0) = \kappa \,
\theta \,\varepsilon_{ij} \,Tr\,\big({\hat \Lambda}^0
(\partial_i {\hat A}_j + {\hat A}_i {\hat A}_j) \big) \approx 0 \,,
\label{eq:g1}
\end{equation}
where ${\hat \Lambda}^0$ is of the same form as $\delta {\hat A}_0$:
\begin{equation}
{\hat \Lambda}^0 {\cal H}_{N-1} \subseteq {\cal H}_{N-1}\,,\quad \quad  
{\hat \Lambda}^0 {\cal H}_{N-1}^\perp = \{0\} \,
\label{eq:l1}
\end{equation}
and where we have changed the sign of (\ref{eq:g1}) 
compared to (\ref{eq:A0}) for future convenience.
In terms of the orthogonal projectors introduced in (\ref{eq:orthgp1}),
\begin{equation}
{\hat \Lambda}^0 = 
{\bf \hat{1}}_{N-1}\, {\hat \Lambda}^0\, {\bf \hat{1}}_{N-1} \,.
\label{eq:orthglp1}
\end{equation}
In the basis spanned by $\{i {\hat P}_{nm} \}$ we have
\begin{equation}
{\hat \Lambda^0} = \sum_{n,m = 0}^{N-2} i 
({\hat \Lambda}^0)_{nm} \,{\hat P}_{nm} \,,
\quad \quad ({\hat \Lambda}^0)_{nm} = 0 \quad  
for \quad  n \quad or \quad m > N-2 \,.
\end{equation}
Anti-hermiticity requires that $({\hat \Lambda}^0)_{nm}^* 
= - ({\hat \Lambda}^0)_{mn}$.

Equation (\ref {eq:g1}) is the statement of Gauss 
law in our matrix model and ${\hat \Lambda}^0$'s 
are non-commutative analogues of the test functions 
of the commutative CS theory.

``Integrating'' by parts $g({\hat \Lambda}^0)$ can be written as
\begin{equation}
g({\hat \Lambda}^0) = \kappa \,\theta \,\varepsilon_{ij} \,Tr\,
\big(\partial_i ({\hat \Lambda}^0 {\hat A}_j) -
\partial_i {\hat \Lambda}^0 {\hat A}_j  + 
{\hat \Lambda}^0 {\hat A}_i {\hat A}_j \big) \approx 0 \,.
\label{eq:g2}
\end{equation}
Here the first term is the trace of a ``total derivative'' 
on a finite-dimensional Hilbert space and it vanishes. Note that
$\partial_i ({\hat \Lambda}^0 {\hat A}_j)$ is identically zero in 
$\{ {\cal H}_{N+\eta+1}^\perp \}$. 
Hence, in our matrix model we can write the Gauss law as 
\begin{equation}
g({\hat \Lambda}^0) = \kappa \,\theta \,\varepsilon_{ij} \,Tr\,\big(-
\partial_i {\hat \Lambda}^0 {\hat A}_j  + {\hat \Lambda}^0 {\hat A}_i 
{\hat A}_j \big) \approx 0 \,.
\label{eq:g3}
\end{equation} 

The conditions on ${\hat \Lambda}^0$ translate to 
those on  $\partial_i \,{\hat \Lambda}^0$ as
\begin{equation}
(\partial_i \,{\hat \Lambda}^0) {\cal H}_N 
\subseteq {\cal H}_N \,,\quad \quad  
(\partial_i \,{\hat \Lambda}^0) {\cal H}_N^\perp = \{0\} \,,
\label{eq:l2}
\end{equation} 
or 
\begin{equation}
(\partial_i{\hat \Lambda}^0) = 
{\bf \hat{1}}_N\, (\partial_i {\hat \Lambda}^0)\, {\bf \hat{1}}_N \,.
\end{equation}

Consider now the quantity
\begin{equation}
q({\hat \Sigma}) = \kappa \,
\theta \,\varepsilon_{ij} \, Tr( -\partial_i {\widehat \Sigma}
{\hat A}_j \,+ {\widehat \Sigma} {\hat A}_i {\hat A}_j) \,,
\label{eq:genexp1}
\end{equation}
for an arbitrary operator ${\widehat \Sigma}$. 
A straightforward calculation shows that
\begin{equation}
\{ q({\widehat \Sigma}_1) \,,  q({\widehat \Sigma}_2) \} 
= -q(\lbrack  {\widehat \Sigma}_1 \,, {\widehat \Sigma}_2 \rbrack) - \kappa \,
\theta \,\varepsilon_{ij} \,
Tr \, {\bf {\hat 1}}_{N+\eta} \, (\partial_i {\widehat \Sigma}_1) \, 
{\bf {\hat 1}}_{N+\eta}\,
(\partial_j {\widehat \Sigma}_2) \,.
\label{eq:genexp2}
\end{equation}
For ${\widehat \Sigma}_i = {\hat \Lambda}_i^0$ $(i=1,2)$ we have
\begin{equation}
q({\hat \Lambda}_i^0) = g({\hat \Lambda}_i^0) \,.
\label{eq:genexp3}
\end{equation}
Also with this substitution the central term in (\ref{eq:genexp2}) becomes
\begin{eqnarray}
-\kappa \,\theta \,\varepsilon_{ij} \,
Tr \, {\bf {\hat 1}}_{N+\eta} \, (\partial_i {\hat \Lambda}^0_1) \, 
{\bf {\hat 1}}_{N+\eta}\,
(\partial_j {\hat \Lambda}^0_2) &=&-\kappa \,
\theta \,\varepsilon_{ij} \, Tr (\partial_i {\hat \Lambda}_1^0) \,
(\partial_j {\hat \Lambda}_2^0) \nonumber \\ 
&=&-\kappa \,\theta \,\varepsilon_{ij} \, Tr \,
\partial_i ({\hat \Lambda}_1^0 \, \partial_j {\hat \Lambda}_2^0) = 0 \,.
\label{eq:genexp4}
\end{eqnarray}
where we have once more made use of the fact 
that the trace of a total derivative term vanishes on a 
finite-dimensional Hilbert space.
Thus for the P.B.'s of $g({\hat {\Lambda}^0})$ we find 
\begin{equation}
\{g({\hat \Lambda}_1^0) , g({\hat \Lambda}_2^0) \} = 
- g \big(\lbrack {\hat \Lambda}_1^0 \,, 
{\hat \Lambda}_2^0 \rbrack \big) \approx 0 \,, 
\label{eq:gauss2}
\end{equation}
assuring that $g({\hat \Lambda}^0)$ are first class constraints.

Consider now ${\widehat \Sigma}={\hat \Lambda} = 
{\hat \Lambda}^\prime + {\hat \Lambda}^0$ for some
${\hat \Lambda}^0$ fulfilling (\ref{eq:l1}) or 
equivalently (\ref{eq:orthglp1}) and
${\hat \Lambda}^\prime$  such that       
\begin{eqnarray}
{\hat \Lambda}^\prime {\cal H}_{N-1} &=& 0 \,,\quad \quad  
{\hat \Lambda}^\prime {\cal H}_{N-1}^\perp 
\subseteq {\cal H}_{N-1}^\perp\,, \nonumber \\
(\partial_i {\hat \Lambda}^\prime) {\cal H}_{N-2} &=& 0\,,\quad \quad  
(\partial_i {\hat \Lambda}^\prime) {\cal H}_{N-2}^\perp \subseteq 
{\cal H}_{N-2}^\perp \,,
\label{eq:l3}
\end{eqnarray}
or
\begin{equation}
0= {\bf \hat{1}}_{N-1}\, {\hat \Lambda}^\prime \, {\bf \hat{1}}_{N-1} \,.
\label{eq:l4}
\end{equation}
In the basis spanned by $\{i {\hat P}_{nm}\}$ we have
\begin{equation}
{\hat \Lambda}^\prime = \sum_{n,m = 
{N-1}}^{\infty} i {\hat {\Lambda}}_{nm}^\prime \,{\hat P}_{nm} \,,
\quad \quad {\hat \Lambda}_{nm}^\prime = 0 \quad for \quad n \quad 
or \quad m < N-1 \,.
\end{equation}
Anti-hermiticity gives $({\hat \Lambda}^\prime)_{nm}^* = 
- ({\hat \Lambda}^\prime)_{mn}$.

Now note that 
\begin{equation}
q({\hat \Lambda}) = 
q({\hat \Lambda}^\prime +{\hat \Lambda}^0) = q({\hat \Lambda}^\prime) 
+ g({\hat \Lambda}^0) \approx q({\hat \Lambda}^\prime) \,.
\end{equation}
Next, from (\ref{eq:l1}) and (\ref{eq:l3}) we observe that
\begin{equation}
\lbrack {\hat \Lambda}^\prime \,, {\hat \Lambda}^0 \rbrack = 0 \,.
\label{eq:com1}
\end{equation}
Hence, we have
\begin{equation}
\lbrack {\hat \Lambda} \,, {\hat \Lambda}_2^0 \rbrack = 
\lbrack {\hat \Lambda}^\prime + {\hat \Lambda}_1^0 \,, 
{\hat \Lambda}_2^0 \rbrack = 
\lbrack {\hat \Lambda}_1^0 \,, {\hat \Lambda}_2^0 \rbrack
\end{equation} 
which is of the form fulfilling (\ref{eq:l1}).

It now follows at once that the P.B.s of 
$q({\hat \Lambda})$ with the Gauss law is
\begin{eqnarray}
\{ q({\hat \Lambda}) \,, 
q({\hat \Lambda}_2^0) \} &=& \{ q({\hat \Lambda}) \,, 
g({\hat \Lambda}_2^0) \}
\nonumber \\
&=&- q(\lbrack {\hat \Lambda}\,, {\hat \Lambda}^0_2 \rbrack) - \kappa \,
\theta \,\varepsilon_{ij} \,
Tr \, {\bf {\hat 1}}_{N+\eta} \, 
(\partial_i {\hat \Lambda}) \, {\bf {\hat 1}}_{N+\eta}\,
(\partial_j {\hat \Lambda}_2^0) \nonumber \\
&=&- q(\lbrack {\hat \Lambda}^\prime\,, {\hat \Lambda}^0_2 \rbrack) - 
q (\lbrack {\hat \Lambda}_1^0  \,, {\hat \Lambda}_2^0 \rbrack)  
- \kappa \,\theta \,\varepsilon_{ij} \,
Tr \, {\bf {\hat 1}}_{N+\eta} \, 
(\partial_i {\hat \Lambda}^\prime) \, {\bf {\hat 1}}_{N+\eta}\,
(\partial_j {\hat \Lambda}_2^0) \nonumber \\
&=&-g (\lbrack {\hat \Lambda}_1^0  \,, 
{\hat \Lambda}_2^0 \rbrack) \approx 0 \,,
\label{eq:chg}
\end{eqnarray} 
where the ``central term'' vanishes. This is 
because $\partial_i {\hat \Lambda}^\prime$ 
is projected to ${\cal H}_{N+\eta}$, and due 
to antisymmetry of the indices it becomes
\begin{equation}
- \kappa \,\theta \,\varepsilon_{ij} \,
Tr \, \partial_j ({\bf {\hat 1}}_{N+\eta} \, 
\partial_i {\hat \Lambda}^\prime \, {\bf {\hat 1}}_{N+\eta}\,
{\hat \Lambda}_2^0)
\end{equation}
which is zero on a finite-dimensional Hilbert space.

In fact one can take the vanishing of the central term as a requirement
on ${\hat \Lambda}^0$'s and find out that the maximally
large subspace of ${\cal H}$ where ${\hat \Lambda}^0$ is 
supported and for which (\ref{eq:com1}) holds is
${\cal H}_{N-1}$.
This explains how we arrive at the particular 
form of the operators for ${\hat A}_0$ and ${\hat \Lambda}^0$
given in (\ref{eq:gaugef1}) and (\ref{eq:l1}) respectively.

As a consequence of (\ref{eq:chg}), $q({\hat \Lambda})$'s 
with ${\hat \Lambda}={\hat \Lambda}^\prime
+{\hat \Lambda}^0$ are first class, and they 
constitute a set of non-abelian observables of our matrix model.

Furthermore, for
\begin{equation}
{\hat \Lambda}_1 = {\hat \Lambda}^\prime+ 
{\hat \Lambda}_1^0 \,, \quad \quad \quad
{\hat \Lambda}_2 = {\hat \Lambda}^\prime+ {\hat \Lambda}_2^0 \,,
\end{equation} 
we have  
\begin{equation} 
q({\hat \Lambda}_1)- q({\hat \Lambda}_2) =  
g({\hat \Lambda}_1 - {\hat \Lambda}_2) \approx 0 \,,
\label{eq:consrv1}
\end{equation}
implying that the actions of $q({\hat \Lambda}_1)$ 
and $q({\hat \Lambda}_2)$ on physical states 
give the same result.
Finally, we compute the P.B.s of $q(\Lambda)$'s and find that 
\begin{equation}
\{ q({\hat \Lambda}_1) \,,  q({\hat \Lambda}_2) \} 
= -q(\lbrack  {\hat \Lambda}_1 \,, {\hat \Lambda}_2 \rbrack) - \kappa \,
\theta \,\varepsilon_{ij} \,
Tr \, {\bf {\hat 1}}_{N+\eta} \, 
(\partial_i {\hat \Lambda}_1) \, {\bf {\hat 1}}_{N+\eta} \,
(\partial_j {\hat \Lambda}_2) \,.
\label{eq:kac4}
\end{equation}

In order to discuss implication of these results we need 
one more ingredient, this is contributed by the P.B.
\begin{equation}
\{({\cal {\hat A}}_l)_{nm} \,, g({\hat \Lambda}^0) \} = 
(\partial_l{\hat \Lambda}^0)_{nm}
- i \lbrack {\hat \Lambda}^0 \,, {\cal {\hat A}}_l \rbrack_{nm} \approx 0 \,,
\end{equation}
for $n \, or \, m \geq N$ and for $n=m=N-1$. 
Thus $({\cal {\hat A}}_l)_{nm}$ for
$n \, or \, m \geq N$ and $({\cal {\hat A}}_i)_{N-1 \, N-1}$ are 
observables of our matrix theory. The algebra of these 
observables is standard and given by (\ref{eq:canonical2})
together with the same restrictions on indices $r$ and $s$ 
as on $n$ and $m$. It is easy to see that
for fixed $N$ and $\eta$ there are $(\eta+1)^2$ of 
these observables. For instance, for $\eta=0$
we have $({\cal A}_i)_{N-1 \, N-1}$ as an observable, 
for $\eta =1$, we have $({\cal {\hat A}}_i)_{N-1 \, N-1}$,
$({\cal {\hat A}}_i)_{NN}\,, ({\cal {\hat A}}_i)_{N \,N-1} \,, 
({\cal {\hat A}}_i)_{N-1 \, N}$ as observables.

From (\ref{eq:kac4}) we see that  the observables $q({\hat \Lambda})$
generate a finite-dimensional Lie algebra with properties similar to a
non-abelian Kac-Moody algebra.  The differences are that this Lie algebra is
finite-dimensional and its central charge is modified, due to the appearance
of the projectors ${\bf {\hat 1}}_{N+\eta}$ in the expression.

Independently of  the value of $\eta$, $q({\hat \Lambda})$ is 
nonzero for nonzero entries of
${\hat \Lambda}_{N-1\, N-1}\,, {\hat \Lambda}_{N-1\, N}$ 
and ${\hat \Lambda}_{N \,N-1}$ in a given ${\hat \Lambda}$.
Thus we have three unique non-abelian Kac-Moody-like 
observables. For fixed $N$ and $\eta$ the rest of such
non-abelian observables are $(\eta+2)^2-1-3=\eta(\eta+4)$ 
in number and they can be constructed
from the $(\eta+1)^2$ observables $({\cal {\hat A}}_i)_{nm}$ given above.

We now investigate the limit $M \rightarrow \infty$.

\section{The Large $M$ Limit}

First we introduce the coherent state $|z>$ by
\begin{equation}
|z> = e^{-\frac{1}{2 
\theta} |z|^2} \sum_{r=0}^\infty \frac{z^r}{\sqrt{\theta^r r!}} \,|r> \,,
\end{equation}
where ${\hat a}|z> = z|z>$, $<z|z> = 1$ 
and $\lbrack {\hat a}\,, {\hat a}^\dagger \rbrack = \theta$.

The diagonal coherent state element of the 
operator ${\hat P}_{M-1 \,M-1} = |M-1><M-1|$ reads
\begin{equation}
{\cal P}_{M-1}(z, {\bar z}) = \frac{1}{\pi 
\theta} <z|M-1><M-1|z> = 
e^{\frac{-|z|^2}{\theta}} \frac{|z|^{2(M-1)}}{\pi \theta^M (M-1)!} 
\label{eq:diagonal1}
\end{equation}
where we have included the normalization factor $\frac{1}{\pi 
\theta}$ so that $\int d^2z {\cal P}(z, {\bar z}) = 1$.
From the definitions
\begin{equation}
{\hat x}_1  = \frac{L}{2 \sqrt{(M-\frac{1}{2}) \,
\theta}}({\hat a}+{\hat a}^\dagger) \,, 
\quad \quad {\hat x}_2  = -\frac{i \sqrt{(M-\frac{1}{2}) \,
\theta}}{L}({\hat a} - {\hat a}^\dagger) \,,
\label{eq:xa1}
\end{equation}
we can deduce the relation between 
$z = (z_1 \,, z_2)$ and $x = (x_1 \,, x_2)$ to be 
\begin{equation}
x_1 = <z|{\hat x}_1|z> = \frac{L}{\sqrt{(M-\frac{1}{2}) 
\theta}} z_1 \,, \quad \quad
x_2 = <z|{\hat x}_2|z> = \frac{2 \sqrt{(M-\frac{1}{2}) \theta}}{L} z_2 \,,
\label{eq:xz}
\end{equation}
or 
\begin{equation}
|z|^2 = z_1^2+z_2^2 = (M-\frac{1}{2}) 
\theta \frac{x_1^2}{L^2}+ \frac{L^2}
{4 {\theta} (M-\frac{1}{2})}x_2^2 \,.\label{eq:z}
\end{equation}
Substituting (\ref{eq:z}) into (\ref{eq:diagonal1}) we 
define the L.H.S. of the resulting expression
by
\begin{equation}
{\cal Q}_{M-1}(x_1 \,,x_2) \equiv {\cal P}_{M-1}(z, {\bar z}) \,.
\end{equation}
From Eqn. (65) we see ${\cal P}_{M-1}$ has  maxima 
\footnote {We thank Fedele Lizzi for his private communications to us 
regarding the nature of the maxima. His comments together with the
discussion in the paper by Lizzi, Vitale and Zampini \cite{lizzi2} have
led us to a better understanding of this issue.}
at the value $z = z_0$ given by $ |z_0|^2 = \theta (M - 1)$.
This implies that the  
function  ${\cal Q}_{M-1}(x_1 \,,x_2)$ has maxima on an ellipse given by
\begin{equation}
\frac{x_1^2}{\Big(\frac{M-1}{M-\frac{1}{2}} \Big) \,L^2}  +
\frac{x_2^2}{\frac{4 \theta^2}{L^2} \,
(M-1)(M-\frac{1}{2})} = 1 \,,
\label{eq:ellipse1}
\end{equation}
%\begin{equation}
%\frac{x_1^2}{\left ( \frac{M-1}{M - \frac{1}{2}} \right ) L^2}
%+ \frac{x_2^2}{\frac{4 \theta^2 (M-1)(M - \frac{1}{2})}{L^2}} = 1 \,,
%\label{eq:ell}
%\end{equation}
the axes of the ellipse being given by
$\sqrt{\frac{M-1}{M-\frac{1}{2}}} L$ and 
$\frac{2}{L} \theta \sqrt{(M-1)(M-\frac{1}{2})}$ respectively. 
Now holding $\theta$ constant and taking $M$ large we see from Figure 1 that
the maxima of ${\cal Q}_{M-1}(x_1 \,,x_2)$ have the geometry of an 
ellipse which is extended along the $x_2$
direction. For large $M$ and fixed $\theta$ and $L$ 
this conclusion is also seen from
the ratio of semimajor axis of the ellipse to its semiminor 
axis, which is approximately
$\frac{2}{L} \, \theta \, M : L$. From this reasoning and from 
Figure 1 and Figure 2
we conclude that the extended elliptical geometry in the limit $M
\rightarrow \infty$ converges to an infinite strip 
along the $x_2$ axis with
peaks at $x_1 = \pm L$. This proves our assertion in 
section 3.2 about the behavior
of diagonal coherent state matrix elements of 
operators on ${\cal H}_M$ for large $M$.
\begin{figure}  
\begin{minipage}[t]{0.45\linewidth}
\centering   
\includegraphics[width=1.0\textwidth, height=0.35\textheight]{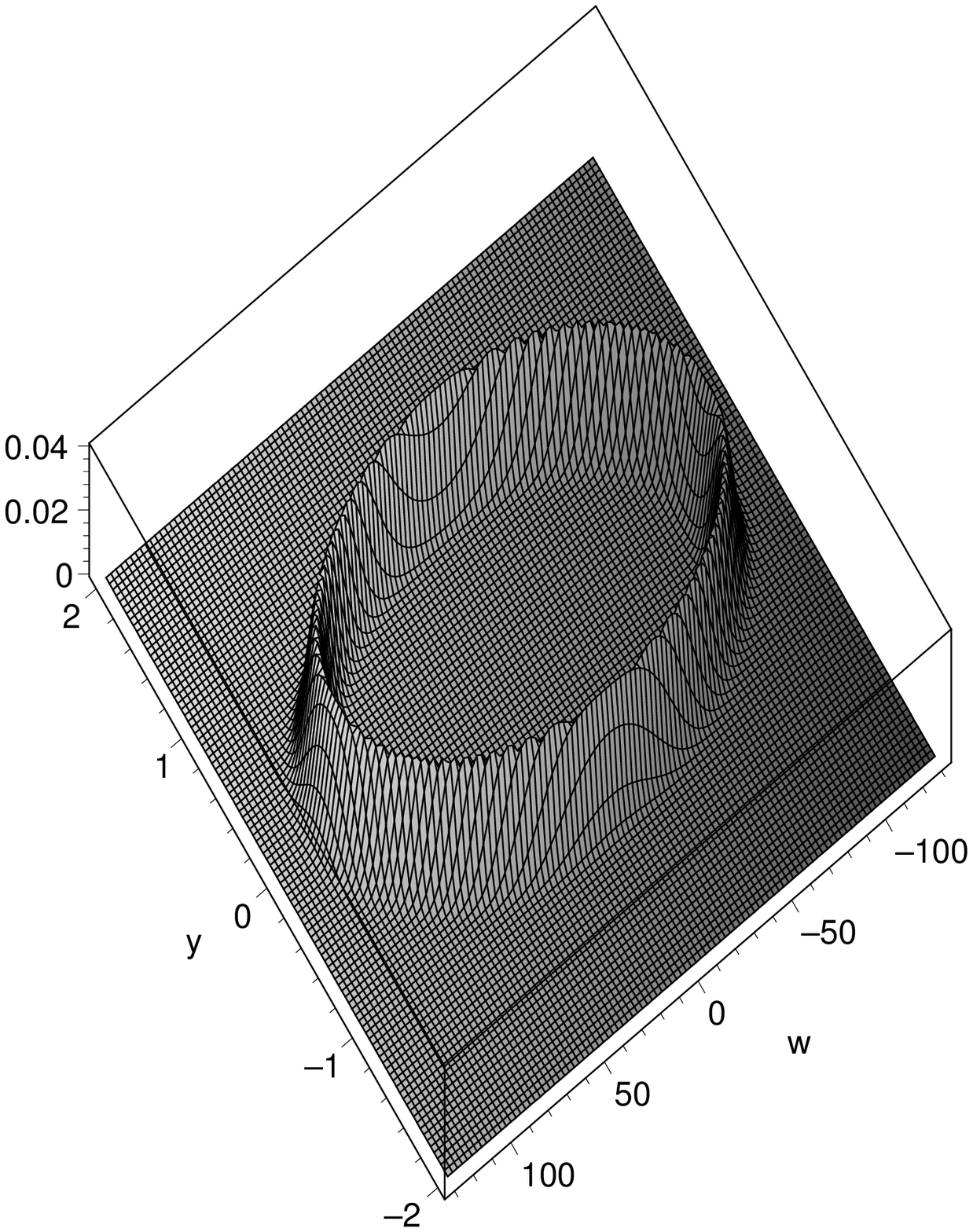}
\caption{{\small Sample plot of ${\cal Q}_{M-1}(x_1 \,,x_2)$ for $M=100$. 
Here the axes are labeled by
$y:=\frac{x_1}{L}$ and $\omega := \frac{L}{2 \theta} x_2$. 
The geometry approximates an infinite strip along
$w$ axis with peaks localized at $y= \pm 1$ as $M \rightarrow \infty$.}}
\end{minipage}%
\hfill
\begin{minipage}[t]{0.45\linewidth}
\centering
\includegraphics[width=1.0\textwidth, height=0.35\textheight]{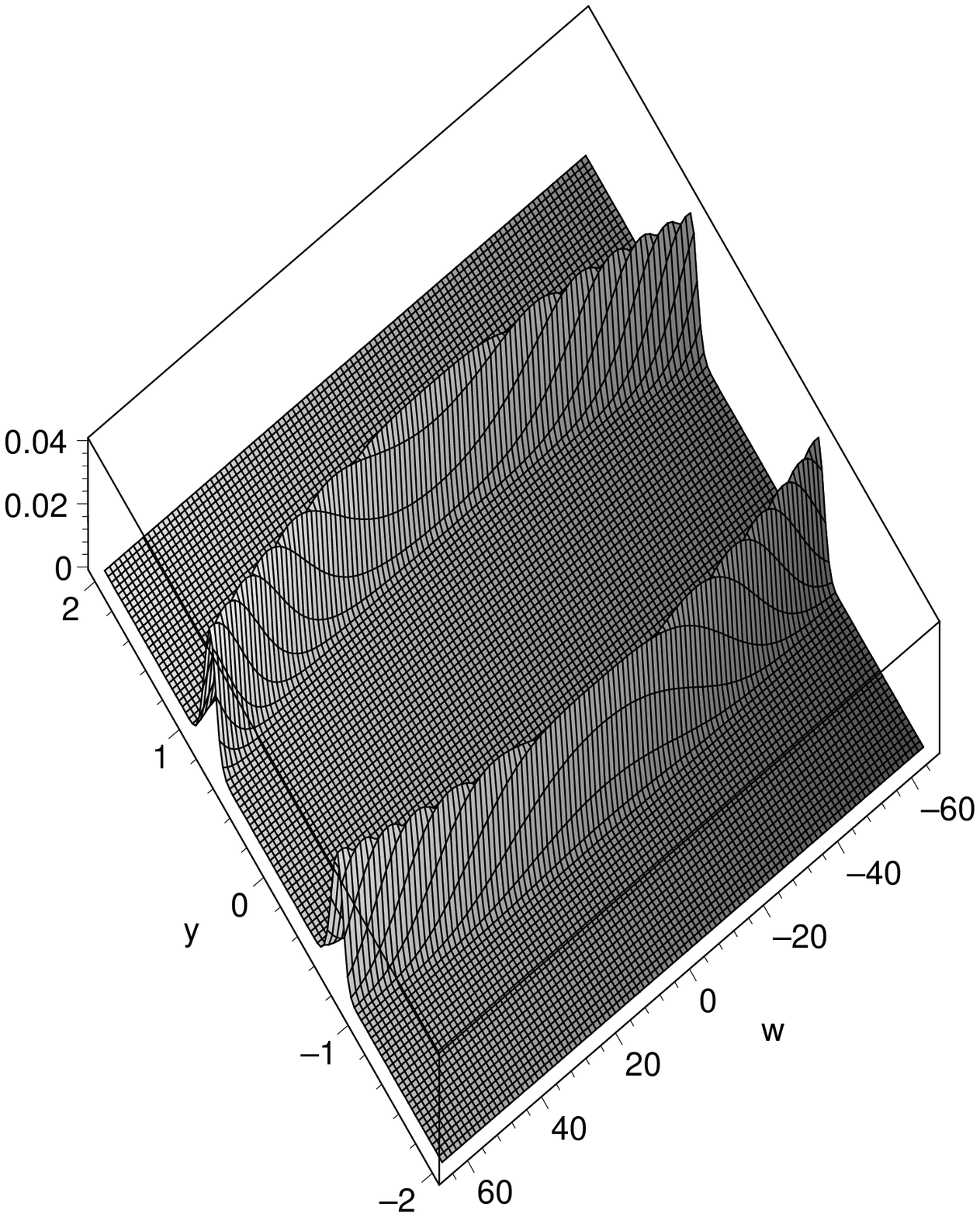}
\caption{{\small Blow up of ${\cal Q}_{M-1}(x_1 \,,x_2)$ for $M=100$
and $|\omega| \leq 65$.}}
\end{minipage}%
\end{figure}
%\begin{figure}
%\begin{center}
%\includegraphics[width=0.6\textwidth, height=0.3\textheight]{elipsoid0}
%\caption{{\small Sample plot of ${\cal Q}_{M-1}(x_1 \,,x_2)$ for $M=100$ 
%and $\omega >0$. Here the axes are labeled by 
%$y:=\frac{x_1}{L}$ and $\omega = \frac{L}{2} x_2$ and $\theta =1$. 
%The geometry approximates an infinite strip along
%$w$ axis with peaks localized at $y= \pm 1$ as $M \rightarrow \infty$.}}
%\end{center}
%\label{fig:Figure 1}
%\end{figure}

In order to get a measure of the sharpness 
of the maxima of ${\cal Q}_{M-1}$, we now compute the width
of the function ${\cal Q}_{M-1}$ at 
$\big( x_1 =  \sqrt{\frac{M-1}{M-\frac{1}{2}}} L \,, x_2=0 \big)$, the
computation for 
$\big( x_1 =  -\sqrt{\frac{M-1}{M-\frac{1}{2}}} L \,, x_2=0 \big)$ being
similar. It may also be noted that
although we do the computation at $x_2=0$ our result is valid for all 
$|x_2|$ independent of $M$ 
in the large $M$ limit, where the geometry is that of an 
infinite strip.
Let the width of the maxima of the function ${\cal Q}_{M-1}$ by denoted by 
$\Delta x_1$. We estimate $\Delta x_1$ by imposing the requirement that 
${\cal Q}_{M-1}(x_1 + \frac{1}{2} \Delta x_1\,,x_2=0) = \frac{1}{e} {\cal
Q}_{M-1}(x_1\,,x_2=0)$. For large values of $M$, the width 
$\Delta x_1$ can be obtained (using Maple) as 
%The point $x_1 \big|_{\frac{1}{e}}$ 
%where ${\cal Q}_{M-1}(x_1\,,x_2=0)$ drops to $\frac{1}{e}$ of its maximum value
\begin{equation}
\Delta x_1 \big|_{M \rightarrow \infty} \rightarrow \sqrt{\frac{2}{M}}\, L + 
{\cal O}(\frac{1}{M}) \,.
\end{equation}
Thus the width $\Delta x_1$ gets narrower as $M$ gets larger.
It is however, important to remark the following: 
The value of the maxima of ${\cal Q}_{M-1}$ for large $M$ is 
proportional to $\frac{1}{\theta \sqrt{M-1}}$. This 
implies that for fixed $\theta$, the height of the peaks decreases
as $M$ gets larger. Nevertheless, the normalization 
integral $\int {\cal Q}_{M-1}(x_1\,,x_2) d^2x =1$ is preserved. A
rough estimate of the volume under the graph of 
${\cal Q}_{M-1}$ would be enough to see that this is the case. 
For large values of $M$, the width 
$\Delta x_1 \approx \frac{1}{\sqrt{M}}$, the height 
$h \approx \frac{1}{\sqrt{M}}$ and $x_2$ extends to order $M$. Thus for
large values of $M$, the volume under the graph is independent of $M$.

It is natural to take the thickness $\Delta \ell$
of the `fuzzy edge'''s in our matrix
model to be given by the 
characteristic width $\Delta x_1$ of ${\cal Q}_{M-1}$.
As $M \rightarrow \infty$ we have $\Delta x_1 =\frac{L}{\sqrt M}
\rightarrow 0$, and consequently we get 
sharp boundaries (i.e. thin edges) at $\pm L$. Thus this limit defines the
non-commutative CS theory on the infinite strip.

We end this section by estimating the spacing between the edge states of our
model. First recall that for a given $\eta \in {\mathbb Z}^+ $, the 
number of edge states in our 
matrix model is $\sqrt{(\eta+2)^2-1} \approx \eta+2$.
Let $\alpha(M)$ be the spacing of these edge states. 
As argued before, the thickness $\Delta \ell$
of the ``fuzzy edge'' is given by $\Delta x_1$. We therefore
have
\begin{equation}
\alpha(M)(\eta+2) = \Delta x_1 \,.
\end{equation}
However, we note that by construction of our model 
we have $0 \leq \eta < M$. This enables us to extract both an upper bound and
a lower bound for the spacing of the edge states $\alpha(M)$. Thus in the
large $M$ limit, using (71) and (72), we find that
\begin{equation}
2\, \alpha(M) \leq \Delta \ell =
\sqrt{\frac{2}{M}} \, L < M \alpha(M) \quad \quad  
\Rightarrow \quad \quad \frac{\sqrt 2 L}{M^{\frac{3}{2}}} < \alpha(M) \leq 
\frac{L}{\sqrt 2 \, M^{\frac{1}{2}}} \,.
\end{equation}

\section{Commutative Limit}

In this section, we would like to 
comment on the commutative limit of our matrix model defined by 
$M \rightarrow \infty$ and $\theta \rightarrow 0$.
First, we define $F(z\,,{\bar z}) := <z|{\hat F}|z>$ 
for an arbitrary operator ${\hat F}$. The 
standard star product is then given by
\begin{equation}
F \star G (z) = <z| {\hat F} {\hat G}|z> \,.
\end{equation}
Making use of the formula 
\begin{equation}
\int d^2 z \,F(z\,, {\bar z}) = 2 \pi \,\theta Tr \,{\hat F} \,,
\end{equation}
we can express (\ref{eq:kac4}) in the 
diagonal coherent state representation as  
\begin{equation}
\{ q(\Lambda_1) \,, q(\Lambda_2) \} 
= -q(\lbrack  \Lambda_1 \,, \Lambda_2 \rbrack_\star) 
-\frac{\kappa}{2\pi} \varepsilon_{ij} \int d^2 z \,{\bf 1}_M \,\star
(\partial_i \Lambda_1) \, \star {\bf 1}_M \, \star \,
(\partial_j \Lambda_2) \,.
\label{eq:cskac}
\end{equation}
The first term in R.H.S. of the above expression vanishes as 
$\theta \rightarrow 0$ since the $\star$-product becomes the 
ordinary product and the commutator of functions $\Lambda_i$ 
under ordinary product is zero. Also in this limit, we see from
(\ref{eq:orthgp1}) that ${\bf {\hat 1}}_M \rightarrow {\bf 1}$. 
Using these in (\ref{eq:cskac}) we recover
the standard $U(1)$ Kac-Moody algebra of the 
edge observables of commutative CS theory:
\begin{equation}
\{ q(\Lambda_1) \,, q(\Lambda_2) \} 
= - \frac{\kappa}{4 \pi} \,\varepsilon_{ij} \, \int d^2x \, 
{\widetilde {(\partial_i \Lambda_1)}}
\, {\widetilde {(\partial_j \Lambda_2)}} \,.
\end{equation}
In getting this final result we have made use of the identity \cite{bal2}
\begin{equation}
\int d^2 z \, F(z,{\bar z} ) = 
\frac{1}{2} \int d^2 x\, {\widetilde F(x_1,x_2)} \,,
\end{equation}
where ${\widetilde F(x_1,x_2)}$ denotes the Moyal representation 
of the operator ${\hat F}$, 
and the fact that the $\star_M$ is removed under 
integration over the Moyal plane.

Finally it may be noted that the parameter $\theta$ may tend to zero in
different fashions resulting to different geometries in the commutative
limit. For example, taking
$\theta \approx \frac{1}{\sqrt{M}}$ still gives a strip geometry in the
large $M$ limit. In this case the height of the maxima of the 
function ${\cal Q}_{M-1}(x_1 \,,x_2)$ at $x_1 =\pm L$ is constant, $x_2$ 
extends to $\pm \sqrt{M}$
and $\Delta x_1$ varies as $\frac{1}{\sqrt{M}}$. To be more precise
for all $\theta M \rightarrow \infty$ as $M \rightarrow \infty$ 
%$\theta > \frac{1}{M}$ the large $M$ limit 
results in a strip geometry. On the other hand, as discussed by
Lizzi, Vitale and Zampini in \cite{lizzi2}, the $\theta \rightarrow 0$ limit 
keeping the product $\theta M$ fixed gives the geometry of a disc.

\section{Concluding Remarks}

In this work we have formulated 
the Chern-Simons theory on an infinite 
strip on the Moyal plane. Our formulation involved
the construction of a new matrix CS model whose 
features are associated with the underlying Moyal plane. Performing canonical
analysis revealed that this matrix model has ``fuzzy'' edges whose 
thickness $\Delta \ell$, we found out to be inversely 
proportional to $\sqrt M$ in the large $M$ limit. Thus 
in this limit our matrix model approximated the non-commutative CS theory
on the infinite strip. Our results show that the edge observables
on the boundaries of the non-commutative infinite strip are given 
by a finite-dimensional
Lie algebra with properties similar to 
that of a non-abelian Kac-Moody algebra.
Our findings generalize the well-known results 
of the usual CS theory to non-commutative manifolds with
boundaries, thereby opening new possibilities for the 
treatment of non-commutative
Quantum Hall Effect and for other applications in such domains.  

\vspace{.5cm}

{\bf Acknowledgments}

\vspace{.5cm}

The authors acknowledge valuable suggestions and comments of Alexander
Pinzul and Allen Stern during various stages of this work. We also thank to
Garnik Alexanian, Brian Dolan, Giorgio Immirzi, Xavier Martin, Denjoe
O'Connor, Peter Pre\v{s}najder and Sachin Vaidya for various useful
discussions and Fedele Lizzi, Patrizia Vitale and Alessandro Zampini for
informing us about their work and for discussions.  The work of A.P.B and
S.K. was supported in part by DOE and NSF under contract numbers
DE-FG02-85ER40231 and INT9908763 respectively. The work of K.S.G. was
carried out partly during the author's visit to the Abdus Salam ICTP,
Trieste, Italy under the Associateship Scheme of ICTP and K.S.G. gratefully
acknowledges the financial support from the Associateship Scheme of ICTP.

\end{document}